\documentstyle[12pt]{article}
\begin{document}

\titlepage
\begin{flushright}
  CERN-TH/96-124\\
  Bielefeld-TP 96/20\\
  JYFL 96-8\\
\end{flushright}

\begin{center}
 {\large\bf Minijet and Transverse Energy Production \\
  in the BFKL Regime }\\[3ex]

{\bf K.J.~Eskola}\footnote{On leave of absence from Laboratory
of High Energy Physics, Department of Physics, 
University of Helsinki, Finland}

CERN, Theory Division, CH-1211 Geneva 23, Switzerland

email: kjeskola@mail.cern.ch\\[2ex]

{\bf A.V.~Leonidov}\footnote{Alexander von Humboldt Fellow, on leave
 from
Lebedev Physics Institute, 117924 Leninski pr. 53, Moscow, Russia}

Theoretische Physik, Fakult\"at f\"ur Physik,\\
Universit\"at Bielefeld, D-33615 Bielefeld, Germany

email: andrei@physik.uni-bielefeld.de\\[2ex]

{\bf P.V.~Ruuskanen}

Department of Physics, University of Jyv\"askyl\"a, \\
P.O. Box 35, FIN-40352 Jyv\"askyl\"a, Finland

email: ruuskanen@jyfl.jyu.fi\\[4ex]

\end{center}

\begin{abstract}
Minijet production in hadronic and nuclear collisions
through a BFKL pomeron ladder is studied for the 
energies of the future LHC heavy-ion collisions.
We use unintegrated gluon densities compatible with
the small-$x$ increase of  parton distributions observed at HERA.
We show that at LHC energies the BFKL minijet and
transverse energy production is at most of the same order
of magnitude as that in the collinear factorization approach.
\end{abstract}
\vfill
CERN-TH/96-124\\
June 1996
 \newpage

Semihard parton scatterings with transverse momenta $p_{\rm T}\sim $
few GeV have been suggested to explain the rapid growth of the inelastic
and total cross sections in high energy $pp(\bar p)$ collisions at
$\sqrt s>20$ GeV \cite{MJET85,XNW91}. Especially in heavy-ion collisions
at very high energies the semihard processes are expected to be
abundant and dominate the transverse energy production in the central
rapidity region \cite{KLL}. Event generators emphasizing the importance
of semihard processes in  ultrarelativistic heavy-ion collisions
at $\sqrt s \ge 200$ $A$GeV have also been actively developed
during recent years \cite{HIJING,PCM}.

Perturbative processes take place at the very early stages of the
time evolution of an ultrarelativistic heavy-ion collision \cite{BM87}.
In the central rapidity region, these semihard  processes occur at time scales
$\tau\sim 1/p_{\rm T}$, {\it i.e.} within the first fractions of fm/$c$,
while the particle production from soft processes is expected to
take longer,  $\sim $ 1 fm/$c$. Therefore, semihard particle
production and associated transverse energy production give
initial conditions for further space-time evolution of the formed
partonic system, which eventually may lead to a thermalized quark-gluon
plasma if the produced system is dense enough. The key feature of
the semihard parton production is that it can be computed by using
perturbative QCD \cite{KLL,BM87}.

At central rapidities the semihard scatterings with momentum
exchanges $p_{\rm T}\sim$ 2 GeV probe typically parton distributions
at $x\sim 2p_{\rm T}/\sqrt s$. Especially in nuclear collisions
at the CERN Large Hadron Collider (LHC) with $\sqrt s= 5.5$ TeV,
these $x$-values fall dominantly in the region of a rapid increase
of the structure function $F_2^p (x,Q^2)$  as observed in deep
inelastic $ep$ collisions at HERA \cite{HERA93}. This increase
persists down to $Q^2=1.5$ GeV$^2$ \cite{HERA96}, thus
strongly enhancing production of semihard partons as discussed in
\cite{EKR}.  Nuclear effects in the parton distributions, in particular
the observed nuclear shadowing of $F_2^A(x,Q^2)$ at small $x$
\cite{NMCE665}, are expected to be important, although the nuclear
gluon distributions are not well known at the moment \cite{KJE93,EQW}.

In the above-mentioned studies, the production mechanism of semihard partons,
minijets, is based on multiple independent $2\rightarrow2$  scatterings
of partons. Collinear factorization is assumed to hold, enabling separation
of the perturbatively calculable hard processes from the
parton distributions containing non-perturbative input. In the leading
twist approximation, the actual hard scattering involves only on-shell
partons. In this paper, we will study a different mechanism for
minijet production, which is not based on collinear factorization.

The small-$x$ increase of $F_2^p$ measured at HERA can be explained
by the Leading Log ($\log Q^2$)  DGLAP-approach \cite{DGLAP}, by the
Double Leading Log ($\log Q^2 \log{1/x}$) approximation \cite{FB},
or by the Leading Log ($\log{1/x}$) BFKL-approach \cite{BFKL,BL78,KMRS}.
In this study, we adopt the BFKL standpoint. Based on this we aim
to study minijet and transverse energy production from a colour singlet
(cut) BFKL ladder spanned between the two colliding nuclei.
We will not make any attempt to include nuclear effects here and,
accordingly,
our conclusions for the minijet cross sections should hold for
$pp$ collisions as well. The basic difference from the previous minijet
studies including collinearly factorized $2\rightarrow2$ on-shell
parton scatterings only \cite{EKR}, is that we now consider
production of semihard gluons emitted by the virtual gluons in the
legs of the BFKL ladder ordered strongly in rapidity, {\it i.e.}
essentially from $2\rightarrow1$ parton processes. This mechanism of
minijet production was first suggested in \cite{GLR,LL94}.
As a new feature, we use the unintegrated gluon distributions
directly normalized to the integrated gluon distributions, which
exhibit a small-$x$ rise compatible with HERA results \cite{AKMS}.
Finally, we compare our results with the more conventional
calculation \cite{EKR} based on collinear factorization.
Our conclusion is that the minijet and transverse energy production
via the BFKL mechanism is at most of the same order of magnitude as the
conventional one.

Minijet production in hadronic collisions from a BFKL ladder in
the case of two tagging (mini)jets separated by a wide rapidity gap
was studied in \cite{MN87,DPT} and more recently in \cite{PK96,RT96}.
A closely related subject in deep inelastic scattering  (DIS) is
the transverse energy flow due to minijets from the BFKL ladder
\cite{ETDIS,LL93}. Conceptually related to our study in this paper
may also be the studies in \cite{McLV}, where collisions of two
virtual gluon clouds are considered in the high energy heavy-ion
collisions. Finally, let us also mention Ref.~\cite{DL94}, where
minijet production by a usual soft pomeron is discussed.

As a starting point for our problem, it is useful to recall
inclusive minijet production in hadron collisions from a (cut)
BFKL ladder as considered
in \cite{MN87} in the case of two tagging jets with a large rapidity
gap in between. The situation on the parton level is illustrated in
Fig.\ 1, where the transverse momenta of the tagging jets are denoted by
${\bf k}_{a{\rm T}},\,{\bf k}_{b{\rm T}}$, and rapidities at the hadron
CMS by $y_a,\,y_b$ with $y_a\gg y_b$. The building blocks of such a ladder
are the reggeized gluon propagators illustrated by the thick lines and
the non-local gauge-invariant effective Lipatov vertices represented
by the black blobs in the legs of the ladder.
Squaring and summing all the $gg\rightarrow (n+2)g$ amplitudes
in the leading $\log(\hat s/\hat t)$ approximation leads then in the
multi-Regge kinematics to a hard pomeron, {\it i.e.} to a strongly
rapidity-ordered perturbative colour singlet gluon ladder, denoted
by $f({\bf q}_{\rm T},{\bf k}_{\rm T},y_a-y_b)$ in Fig.~1.
Note that in the multi-Regge kinematics, only the transverse
momenta of the legs of the ladder become important, so that
$q^2\approx -{\bf q}_{\rm T}^2\equiv -q_{\rm T}^2$. Also, when
intrinsic transverse momenta of the incoming partons are neglected, we have
${\bf q}_{\rm T} = {\bf -k}_{a{\rm T}}$ and
${\bf k}_{\rm T} = {\bf k}_{b{\rm T}}$.
An addition of a rung to the pomeron ladder is governed by
an inhomogeneous BFKL equation \cite{BFKL,BL78}. For  details,
we refer to the lecture notes of Del Duca \cite{DELDUCA},
here we merely cite the result for the cross section,
\begin{equation}
\frac{d\sigma}{d^2{\bf k}_{a{\rm T}} d^2{\bf k}_{b{\rm T}} dy_a dy_b} =
x_ag(x_a,\mu^2)\,x_bg(x_b,\mu^2) \,
\frac{d\hat\Sigma}{d^2{\bf k}_{a{\rm T}} d^2{\bf k}_{b{\rm T}} dy_a dy_b}
\end{equation}
where
\begin{equation}
\frac{d\hat\Sigma}{d^2{\bf k}_{a{\rm T}} d^2{\bf k}_{b{\rm T}} dy_a dy_b} =
\frac{4N_{\rm c}^2\alpha_s^2}{N_{\rm c}^2-1}   \,
\frac{1}{k_{a{\rm T}}^2}     \,
2f({\bf q}_{\rm T}, {\bf k}_{\rm T}, y_a-y_b)  \,
\frac{1}{k_{b{\rm T}}^2}.
\label{1LADDER}
\end{equation}
In the first approximation, only gluons can be considered,
with densities $g(x,\mu^2)$.  The factorization scale is denoted by $\mu^2$.
Fractional momenta of the incoming gluons are  $x_a$ and $x_b$, which
in the multi-Regge kinematics become
\begin{equation}
x_a \approx \frac{k_{a{\rm T}}}{\sqrt s}{\,\rm e}^{\,y_a},
\,\,\,\,\,\,\,\,\,\,\,\,\,\,\,\,\,\,\,\,
x_b \approx \frac{k_{b{\rm T}}}{\sqrt s}{\,\rm e}^{\,-y_b}.
\end{equation}

With the tagging jets the calculation is well defined, since
large transverse momenta are required for the tagging jets
so that the coupling of the
ladder to the jets becomes perturbative. In this case one can
still rely on collinear factorization and use the integrated parton
densities $xg(x,\mu^2)$.
Another important point in this calculation is that when all the
radiative and virtual corrections are neglected, one is left with
the inhomogeneous term of the BFKL equation only. At this limit the
ladder shrinks into $2f({\bf q}_{\rm T},{\bf k}_{\rm T},y)\rightarrow
\delta^{(2)}({\bf q}_{\rm T}-{\bf k}_{\rm T})$,
and the Born limit for the two jets separated by a large rapidity
interval is recovered \cite{DELDUCA}.

Within the framework of tagging jets, also more exclusive minijet
processes have been studied by drawing a minijet out of the ladder
\cite{DPT}. It is straightforward to show that after pulling out
a minijet, the gluon emissions before and after the chosen minijet
can be summed, resulting as a pomeron ladder on each side of the pulled
minijet, as illustrated in Fig.\ 2. The cross section at the parton level
becomes \cite{DPT}

\begin{eqnarray}
\frac{d\hat\Sigma} {d^2{\bf k}_{a{\rm T}} d^2{\bf k}_{b{\rm T}}
d^2{\bf k}_{c{\rm T}} dy_a dy_b dy_c}  =
\frac{4N_{\rm c}^2\alpha_s^2}{N_{\rm c}^2-1} \,
\frac{\alpha_s N_{\rm c}}{\pi^2}       \,
\frac{1}{k_{c{\rm T}}^2}            \,
\int d^2{\bf q}_{1{\rm T}}d^2{\bf q}_{2{\rm T}}\cdot \nonumber
\\\nonumber
\\
\cdot\,\delta^{(2)}({\bf k}_{c{\rm T}}-
{\bf q}_{1{\rm T}}+{\bf q}_{2{\rm T}})\,
\frac{2f( {\bf k}_{a{\rm T}}, {\bf q}_{1{\rm T}}, y_a-y_c)}{k_{a{\rm T}}^2} \,
\frac{2f( {\bf q}_{2{\rm T}}, {\bf k}_{b{\rm T}}, y_c-y_b)}{k_{b{\rm T}}^2},
\label{2LADDER}
\end{eqnarray}
where the minijet drawn from the middle of the pomeron ladder is labelled
as $c$. The factor $\alpha_sN_{\rm c}/(\pi^2 k_{c{\rm T}}^2)$
is a combination of the phase-space factor and the two
Lipatov vertices associated with the step $c$.
The factors 2 are related to the  normalization of the
ladder $f$ as in Eq.\ (\ref{1LADDER}). Again, if the radiative and virtual
corrections in the ladder are neglected, one recovers the Born limit
for production of three jets well separated in rapidity  \cite{DELDUCA}.

What we do in the following is that we simply relax the requirement of
having tagging jets, since we want to study the leading BFKL minijet
production mechanism, which is $\sim\alpha_s$. This means, unfortunately,
that coupling of the pomeron ladder to the hadron becomes essentially
soft in $q_{\rm T}$ and $k_{\rm T}$, {\it i.e.} non-perturbative.
This in turn results in the fact that while the other leg of
the ladder is coupled to one constituent of the incoming hadron, in the
squared graph the other leg may be coupled to another constituent, as shown
in Fig.\ 3. This is nicely illustrated in a classic paper \cite{BL78}
where photon-photon scattering via heavy quarks was studied.
Also, now that we do not require any tagging jets
we have to give up collinear factorization, and,
for the forward scattering amplitude, we do not have a perturbative
Born limit to compare with, either. Therefore, the best we can do
is to adopt the procedure for deep inelastic scattering in \cite{AKMS},
where an addition of each rung into the pomeron ladder between the two
hadrons or nuclei is expected to be described by the {\it homogeneous}
BFKL equation. Justification for using a homogeneous evolution
equation is that the $x$-dependence of the hadron form factor describing
coupling of the gluon ladder to the hadron
is negligible  with respect to the $x$-dependence
generated by the BFKL evolution \cite{BL78}.

The homogeneous BFKL equation for the unintegrated gluon densities
$f(x,q_{\rm T}^2)$ given in \cite{AKMS}, which we will utilize in 
our problem, can be derived from the homogeneous BFKL equation for
$f({\bf q}_{\rm T},{\bf k}_{\rm T},y)$ in  \cite{BFKL,MN87}
by first integrating over the last momentum exchange ${\bf k}_{\rm T}$
in the pomeron ladder, then scaling with $q_{\rm T}^2$ and integrating
over the azimuthal angle in the kernel, setting $y = \log(1/x)$ and
finally scaling with $x$. The BFKL equation then becomes \cite{AKMS}
\begin{equation}
-x\frac{\partial f(x,q_{\rm T}^2)}{\partial x} =
\frac{\alpha_s N_{\rm c}}{\pi}q_{\rm T}^2
\int_{0}^{\infty}
\frac{dq_{1{\rm T}}^2}{q_{1{\rm T}}^2}
\bigg[
\frac{f(x,q_{1{\rm T}}^2)-f(x,q_{\rm T}^2)}{|q_{\rm T}^2-q_{1{\rm T}}^2|} +
\frac{f(x,q_{\rm T}^2)}{\sqrt{q_{\rm T}^4+4q_{1{\rm T}}^4}}
\bigg].
\label{HBFKL}
\end{equation}

The above equation is scale-invariant, so that additional information
for fixing the normalization of the $f(x,q_{\rm T}^2)$ is needed. The
integrated gluon densities of a proton, determined through the global
analysis of parton distributions  \cite{AKMS}, will provide us with
this input:
\begin{equation}
xg(x,Q^2) = \int^{Q^2}\frac{dq_{\rm T}^2}{q_{\rm T}^2}f(x,q_{\rm T}^2).
\end{equation}
It should be noted that the non-perturbative nature of the
pomeron-hadron coupling is now hidden into the initial
$q_{\rm T}^2$-distribution $f(x_0,q_{\rm T}^2)$, which must be
supplied at fixed  $x_0\le 0.01$.
Equation (\ref{HBFKL}) can then be solved numerically to obtain $f(x,q_{\rm
T}^2)$. In our discussion we will return to the treatment of the 
infrared problems of Eq.\ (\ref{HBFKL}) but here will rely on the 
analysis of \cite{AKMS}.

Now we are ready to write down the formula for inclusive minijet production
from the BFKL pomeron ladder between the two colliding hadrons (nuclei)
as illustrated in Fig.\ 3.
Based on Eq.\ (\ref{2LADDER}), we get the factor
$\alpha_sN_{\rm c}/(\pi^2 p_{\rm T}^2)$ when pulling out a minijet by
fixing the momentum of one rung of the pomeron ladder. When doing this we
form a new ladder on each side of the
minijet. The coupling of these ladders to the hadrons or nuclei
is contained in the initial condition for the unintegrated gluon
distribution $f(x,q^2)$, as explained above. Then, the cross section for
inclusive minijet production is bound to have the same structure as in
Eq.\ (\ref{2LADDER}), and one obtains \cite{GLR,LL94}
\begin{equation}
\frac{d\sigma^{\rm jet}}{d^2{\bf p}_{\rm T}dy} =
K_N
\frac{\alpha_sN_{\rm c}}{\pi^2} \,
\frac{1}{p_{\rm T}^2} \,
\int d^2{\bf q}_{1{\rm T}}d^2{\bf q}_{2{\rm T}} \,
\delta^{(2)}({\bf p}_{\rm {\rm T}}-{\bf q}_{1{\rm T}}+{\bf q}_{2{\rm T}}) \,
\frac{f(x_1,q_1^2)}{q_{1{\rm T}}^2} \,
\frac{f(x_2,q_2^2)}{q_{2{\rm T}}^2}
\label{MINIJET}
\end{equation}
where $p_{\rm T}$ and $y$ are the transverse momentum and the rapidity
(in the hadron CMS) of the minijet. From  momentum
conservation and multi-Regge kinematics the momentum fractions become
\begin{equation}
x_1 \approx \frac{p_{\,\rm T}}{\sqrt s}{\,\rm e}^{\,y},
\,\,\,\,\,\, \,\,\,\,\,
x_2 \approx  \frac{p_{\,\rm T}}{\sqrt s}{\,\rm e}^{-y}.
\label{x1x2}
\end{equation}
Due to the fact that in this case we do not have an ``external'' hard
probe like the virtual photon with an associated quark box as
in DIS, nor an on-shell Born cross section to relax into, we cannot
determine the overall dimensionless normalization constant $K_N$ exactly.
However, we are able to fix the {\it slope} of the minijet
$p_{\rm T}$-distribution. Then, based on this slope we will
discuss the  upper bound for $K_N$ and actually argue that 
$K_N\,{\buildrel < \over {_\sim}}\,1$. It is possible that an analysis 
of two-jet emission will clarify this issue. This demands in fact 
considering a $k_{\rm T}$-factorized form of the two-jet production 
cross section \cite{CCH}. By comparing this rate with the one
evaluated directly from the BFKL ladder one would hope to get an 
absolute normalization to the off-shell gluon flux.

Instead of numerically solving the homogeneous BFKL equation
(\ref{HBFKL}), we will use a simple parametrization for the solution,
motivated by the form of the solution in the limit of asymptotically
small $x$ \cite{BL78,AKMS}:
\begin{equation}
f(x, q_{\rm T}^2) =
\frac{C}{x^{\lambda}} \,
\bigg({\frac{q_{\rm T}^2}{q_0^2}} \bigg)^{\frac{1}{2}}\,
\frac{{\bar{\varphi}}_0 }{\sqrt{2\pi\lambda''\ln(1/x)}}
\exp\bigg[
-\frac{\ln^2(q_{\rm T}^2/\bar q^2)}{r2\lambda''\ln(1/x)}
\bigg]\,.
\label{FIT}
\end{equation}
In this expression $\lambda = 4 {\bar{\alpha}}_s \ln 2$, $\lambda'' =
28 {\bar{\alpha}}_s \zeta (3)$, ${\bar{\alpha}}_s=3 \alpha_s/\pi$,
$\zeta(3) = 1.202$ is the Riemann zeta function and $\bar\varphi_0$, $q_0$
and $\bar q$ are parameters characterizing the initial distribution
in \cite{AKMS}.
We reproduce the width of the solution with an additional
parameter $r$.
For definiteness we shall take $\alpha_s = 0.2$, resulting in
the same slope ($\sim x^{-0.5}$) as in the MRSD-' set of parton
distributions \cite{MRSD-'}. By choosing $\bar q =q_0=1$,
$C \bar\varphi = 1.19$ and $r=0.15$, the proposed parametrization
(\ref{FIT}) describes with a satisfactory accuracy the gluon 
distribution, which is obtained as the numerical solution of the BFKL
equation in the global analysis \cite{AKMS}\footnote{We
are grateful to  D.M.~Ostrovsky for a numerical check
of this statement.}.
For comparison with Ref.\ \cite{AKMS},  we show the
unintegrated gluon distributions of Eq.\ (\ref{FIT}) in Fig.\ 4 as
functions of $q_{\rm T}^2$ at different values of $x$.

Proceeding then to the computation of the  minijet cross sections, we
integrate over the azimuthal angle between ${\bf p}_{\rm T}$ and
${\bf q}_1$ in Eq.\ (\ref{MINIJET}) and fix $K_N = 1$ and $y=0$.
Results for the minijet cross sections $d\sigma^{\rm jet}/(dp_{\rm T}dy)$
at the LHC heavy-ion energy 
$\sqrt s = 5.5$ TeV and at the UA1
energy $\sqrt s = 900$ GeV are shown in Figs. 5a and b. These are
the main results of this paper. The corresponding cross sections
from the more conventional, collinearly factorized leading twist
lowest order (CFLTLO) $2\rightarrow2$ mechanism (see  \cite{EKR})
with the MRSD-' parton distributions,  are also shown for
comparison. In the MRSD-' parton distribution functions we have made
a scale choice $Q=p_{\rm T}$, so we can only proceed down to
$p_{\rm T} =\sqrt 5$ GeV, the lowest scale for this set.
With the BFKL mechanism, we extend our computation down to 1 GeV.
Note also that no $K$-factor to simulate the
next-to-leading order (NLO) terms is used for the CFLTLO
computation.

In the BFKL approach only gluons are considered, and we have fixed
$\alpha_s = 0.2$ in Eq.\ (\ref{MINIJET}). In order to compare the two
production mechanisms at the same level of approximation, we also plot
the corresponding curves for the collinearly factorized results.
However, since in the analysis of $f(x,q_{\rm T}^2)$ in \cite{AKMS},
the strong coupling constant was set to run --- phenomenologically,
because NLO corrections to the BFKL ladder are not yet known ---
we show also the BFKL result with $\alpha_s(p_{\rm T})$.

Let us then come to the question of the normalization constant $K_N$.
We expect the BFKL mechanism to be valid and potentially important
only at $x<0.01$, where the rapid rise of the structure function
$F_2(x,Q^2)$ is observed \cite{HERA93,HERA96}. Based on
Eq.\ (\ref{x1x2}), the region of validity would then be
$p_{\rm T}<9$ GeV  for $\sqrt s = 900$ GeV at $y=0$ in Fig.\ 5b.
On the other hand, the experimentally measured jet cross sections
at $p_{\rm T}>5.5$ GeV \cite{UA188} can be explained through
the collinear factorization mechanism for jet production rather 
than through the BFKL mechanism considered here. When the next-to-leading 
order corrections to CFLTLO jet cross section are included,
the measured jet cross sections are well reproduced \cite{EKS,EWatHPC}.
Let us also note that with a scale choice $Q=p_{\rm T}$ the NLO
calculation can be reproduced by multiplying the Born level cross
section by a factor between 2 and 1.5, depending on $\sqrt s$ and on
the size of the jet \cite{EWatHPC}. This tells us that there is
no room for an {\it additional} jet production mechanism of the
same order of magnitude. Given this, and comparing the two mechanisms
at the same level of approximation, the conclusion from Fig.\ 5b is
that the BFKL mechanism in Eq.\ (\ref{MINIJET}) can indeed at most
have $K_N\sim 1$. In fact, this  provides an upper bound
for the non-leading BFKL corrections as well.

Up to this point, everything we have considered applies to $pp$
collisions. The first estimate, without nuclear modifications
to the parton densities, of the average number distributions
of produced semihard gluons in central $AA$ collisions can be obtained
simply by multiplying the results in Fig.\ 5 by the nuclear overlap
function, $T_{AA}({\bf b=0})\approx A^2/(\pi R_A^2)$ \cite{KLL}.
For central Pb-Pb collisions, $T_{\rm PbPb}({\bf 0}) = 32/{\rm mb}$.

As the last step, we estimate the transverse energy production due to
the BFKL minijets in nuclear collisions. From the point of view of
quark-gluon-plasma formation, and of its further evolution, we are
mostly interested in the transverse energy deposit into the central
rapidity region $y\sim 0$. An estimate for the rapidity
distribution of the average transverse energy in an $AA$ collision at
an impact parameter ${\bf b=0}$ is obtained as

\begin{equation}
\frac{d\bar E_{\rm T}^{AA}}{dy}{\bigg|}_{y=0} =
T_{AA}({\bf b})
\int_{p_0} dp_{\rm T}
p_{\rm T} \frac{d\sigma^{\rm jet}}{dp_{\rm T}dy}{\bigg|}_{y=0}.
\label{ET}
\end{equation}
Since the  minijet cross sections computed for Fig.\ 5 are approximately
constant at rapidities $|y|<0.5$, Eq.\ (\ref{ET}) also gives a good
estimate for the transverse energy deposit into the central unit of
rapidity. By using the BFKL cross sections in Fig.\ 5 with $p_0 = 1$ GeV
we obtain the results shown in Table 1 for the  LHC heavy-ion energy
$\sqrt s=5.5 \,A$TeV. Again, comparison with the CFLTLO results
with $p_0 = \sqrt5$ GeV is made in this table. We remind the reader 
that the numbers quoted for the BFKL minijets are with $K_N = 1$ and
we do not make any attempt to include nuclear modifications 
\cite{KJE93,EQW} in any of the gluon distributions considered.

\begin{table}
\begin{center}
\begin{tabular}{|c|c|c|}\hline
&&\\
$d\bar E_{\rm T}^{\rm PbPb}/dy$ &BFKL & CFLTLO\\
$y=0$, ${\bf b}=0$  &$K_N=1$, $p_0 = 1$ GeV &glue only, $p_0 = \sqrt 5$ GeV \\ \hline
&&\\
$\alpha_s = 0.2$                & 3060 GeV              & 3060 GeV   \\
&&\\ \hline
&&\\
$\alpha_s = \alpha_s(p_{\rm T})$      & 4940 GeV              & 4870 GeV    \\
&&\\ \hline
\end{tabular}
\end{center}
\caption{The average transverse energy at $y=0$ carried by minijets with $p_{\rm T}\ge p_0$ in central Pb-Pb collisions at $\protect\sqrt s =5.5$ $A$TeV.}
\end{table}

Concerning our results, we would like to point out the
following observations:

1. Although the BFKL mechanism we have considered here is leading in
powers of $\alpha_s$
$-$ BFKL $\sim \alpha_s$ and CFLTLO $\sim \alpha_s^2$ $-$
and although the transverse momenta in the
BFKL ladder have enlarged phase-space in the sense that they are
not ordered as in the DGLAP-ladder, it is not enough to overcome
the CFLTLO contribution at $p_{\rm T} \ge p_0 \sim 2$ GeV.

2. The slopes of the BFKL and CFLTLO computations in Fig.\ 5 are quite
similar at large $p_{\rm T}$. This demonstrates that at fixed $\sqrt s$
the relevant physical scale is the transverse momentum $p_{\rm T}$.
Since the gluon distributions have the slope $\sim x^{-0.5}$
and since $x\sim p_{\rm T}/\sqrt s$, the jet cross sections scale
as  $d\sigma^{\rm jet}/dp_{\rm T}\sim (x_1x_2)^{-0.5}/p_{\rm T}^3
\sim \sqrt s/p_{\rm T}^4$ for both mechanisms.

Let us next discuss the potential caveats and uncertainties.
As discussed above, due to the non-perturbative coupling of the
pomeron to the hadrons or nuclei, and lacking the corresponding
partonic Born process as a special limiting case, we were
able to fix the slope of the minijet production only, not the
absolute magnitude. Based then on the measured jet cross sections 
\cite{UA188} and the collinearly factorized calculations, we argued 
that the production of minijets through the BFKL mechanism can at 
most be compatible with minijet production in the collinear 
factorization approach \cite{EKR}.

However, both approaches have problems when $p_{\rm T}\rightarrow0$,
which we did not discuss above. At this limit the  CFLTLO cross sections
grow rapidly, and, especially with the HERA parton distributions,
it is not plausible to go much below $p_{\rm T}=p_0\sim2$ GeV since
this leads to an overprediction of the measured charged hadron distributions
\cite{UA190}.

On the other hand, also the BFKL approach has serious infrared problems. 
As seen in Eq.\ (\ref{HBFKL}), a contribution to the BFKL evolution comes 
from the soft region $q_{\rm T}<q_{\rm T}^0 \sim$ 1 GeV, the importance of
which is enhanced if a running coupling is used. The infrared region
was treated in \cite{AKMS} by introducing an $x$-independent form factor 
at $q_{\rm T}<q_{\rm T}^0=1$ GeV together with $\alpha_s(q_{\rm T}^2)$. 
For a more detailed discussion, we refer to \cite{AKMS} and references 
therein. We believe that in spite of this phenomenological input in the 
soft sector, the correct order of magnitude of the BFKL results can be 
obtained since, first, the high-$q_{\rm T}^2$ region can be computed 
perturbatively from the BFKL equation and, second, the normalization 
of the unintegrated gluon distributions to the integrated ones can be made.
We would also expect that if the BFKL mechanism is indeed the reason
for the rapid rise of the HERA parton densities at small $x$, it could
dominate the minijet production in the region of $p_{\rm T}<p_0$, where the
CFLTLO calculation does not apply.

The subleading contributions are not yet known for the
BFKL approach, although the work is currently in progress
\cite{DUCA96}. For the $2\rightarrow2$ cross sections
of massless partons, the NLO terms have already been known for a while
\cite{ES}, and they have been applied to the case of jet production
\cite{EKS}. However, one should keep in mind that minijet production
is conceptually different from the jet production since minijets
are not observed as individual jets. For minijets the NLO analysis
can be made for infrared-safe quantities, such as the transverse
energy deposit in $|y|\leq 0.5$. This task has not been completed yet.
Especially, it remains to be seen what the scale dependence of the
transverse energy deposit turns out to be.

The last potential caveat, well hidden in our BFKL
approach\footnote{We thank Yu. Dokshitzer for pointing this out.},
is related to the inclusive and more exclusive processes.
In \cite{MN87}, the calculation is well defined, since
the minijet production considered is fully inclusive. One finds
a complete cancellation of infrared singularities between the
virtual and real emissions in the pomeron ladder, leading to
the explicit form of the kernel in Eq.\ (\ref{HBFKL}).
When the coherence of the ladder is broken by extracting a
minijet, a potential problem with the exclusiveness
and the cancellation of the singularities arises (see \cite{Marc}).
At the moment we do not know precisely how to address this problem.
However, since we are computing semi-inclusive quantities, 
in particular the transverse energy flow with less 
infrared-sensitivity, we believe that our estimates of the 
magnitude of minijet production with $p_T\ge 1$~GeV and of the 
associated  $\bar E_T^{AA}$ in $|y|\leq 0.5$ are reliable.

Finally, we would like to discuss the connection to other studies.
Recently, there has been a lot of activity in developing an approach 
to semihard parton production in ultrarelativistic heavy-ion collisions, 
based on an idea of colour charge distributions of the nuclear valence 
quarks moving along the light cone and creating clouds of virtual gluons 
around them \cite{McLV}. In this approach, production of semihard quanta 
will result from a collision of two such incoming clouds. The leading 
mechanism of minijet production would then obviously be through
$2\rightarrow1$ processes, {\it i.e.} of order $\alpha_s$, not
$\alpha_s^2$ as in the collinearly factorized mechanism. Conceptually, 
this may be similar to the BFKL mechanism we have considered here, 
although, to our knowledge, an exact relation is not known at the moment.

To summarize, we have computed minijet production from a BFKL 
pom\-eron ladder and estimated the transverse energy production 
in the mid-rapidity in a central Pb-Pb collision at 
$\sqrt s=5.5$ $A$TeV. We have used unintegrated gluon densities 
similar to those of Ref.\ \cite{AKMS}, so as to show that the BFKL 
contribution is  at most of the same order of magnitude as
the collinearly factorized  one in the considered region.
However, with the BFKL mechanism one could be able to bridge the 
way into the region $p_{\rm T}< p_0$ in a manner at least partly 
controlled by perturbative QCD.
\bigskip

{\bf Acknowledgements:}
We are grateful to Yu.~Dokshitzer, K.~Kajantie, E.~Laenen, L.~McLerran
and H.~Satz for useful discussions. We would also like to thank Nordita 
for hospitality during part of this work and the Finnish Academy
(K.J.E.) for financial support.  The work of A.V.L. was partially supported 
by the Russian Fund for Basic Research, Grant 93-02-3815.

\newpage
\centerline{\large\bf Figure Captions}
\bigskip
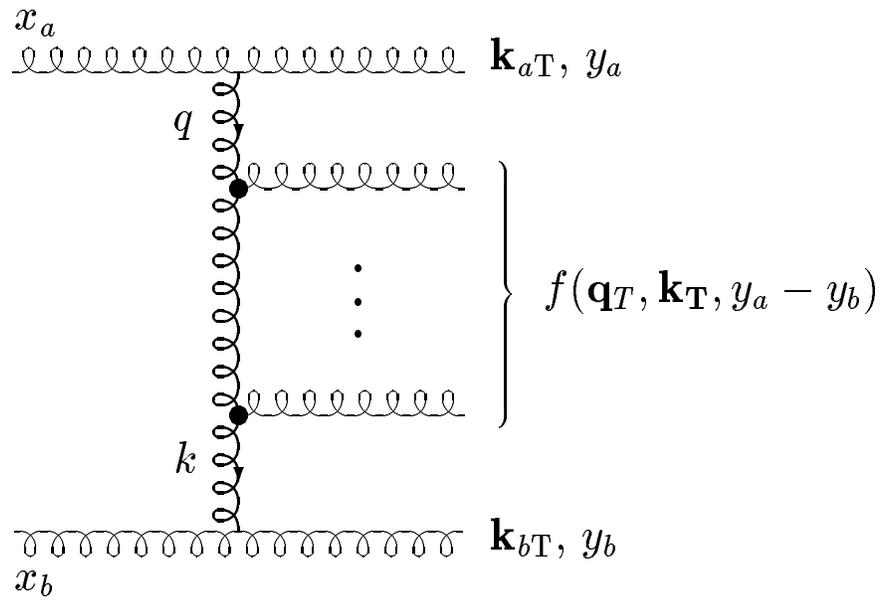
\noindent {\bf Fig.\ 1.}
Fully inclusive minijet production with tagging jets $a$
and $b$ \cite{MN87}. The BFKL pomeron ladder is denoted by
$f({\bf q}_{\rm T},{\bf k}_{\rm T}, y_a-y_b)$, and only the cut
ladder is shown. The thick lines represent the exponentiated
gluon propagators and the black blobs stand for the effective
non-local Lipatov vertices.
\bigskip

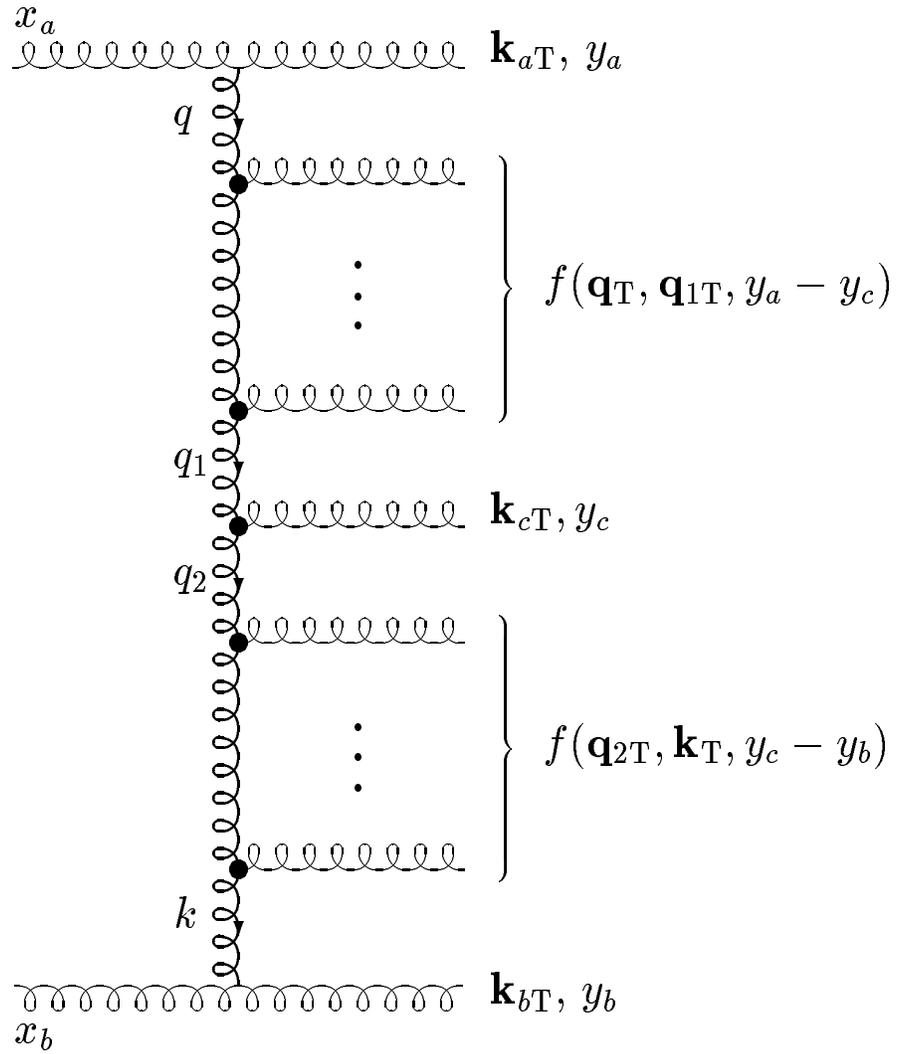
\noindent {\bf Fig.\ 2.}
Production of minijet $c$ from the cut BFKL pomeron ladder of Fig.~1
\cite{DPT}. A new ladder arises on each side of the minijet.
\bigskip

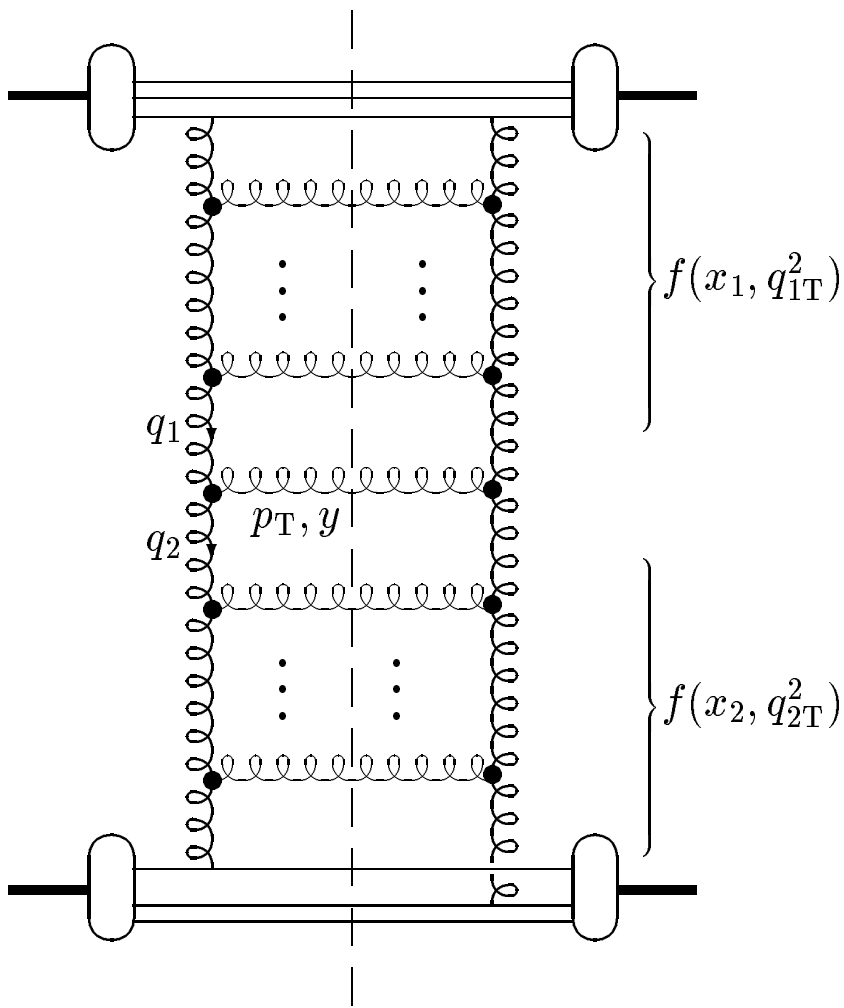
\noindent {\bf Fig.\ 3.}
Production of minijets from the cut BFKL pomeron ladder without
tagging jets. The ovals with one thick line and three thin ones
attached represent the incoming protons, and
$f(x_{1(2)},q_{1(2){\rm T}}^2)$ describe the BFKL pomeron ladders
obeying Eq.\ (\ref{HBFKL}). For details, see the text. The cross 
section for minijet production through this mechanism is given
in Eq.\ (\ref{MINIJET}).
\bigskip

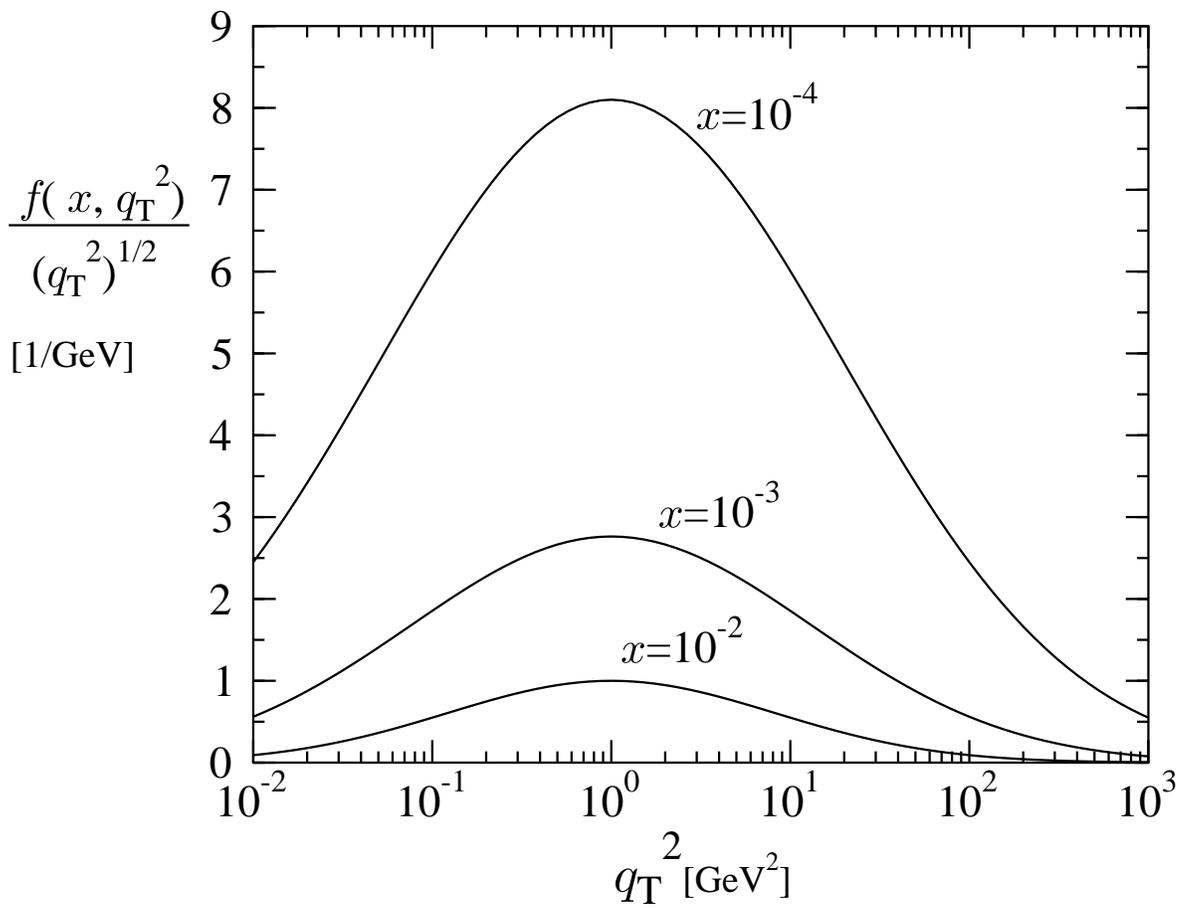
\noindent {\bf Fig.\ 4.}
The unintegrated gluon distributions $f(x,q_{\rm T}^2)$ divided by
$\sqrt {q_{\rm T}^2}$ as given by the parametrization in
Eq.\ (\ref{FIT}) as a function of $q_{\rm T}^2$. The curves 
correspond to three different, fixed values of $x$.
\bigskip

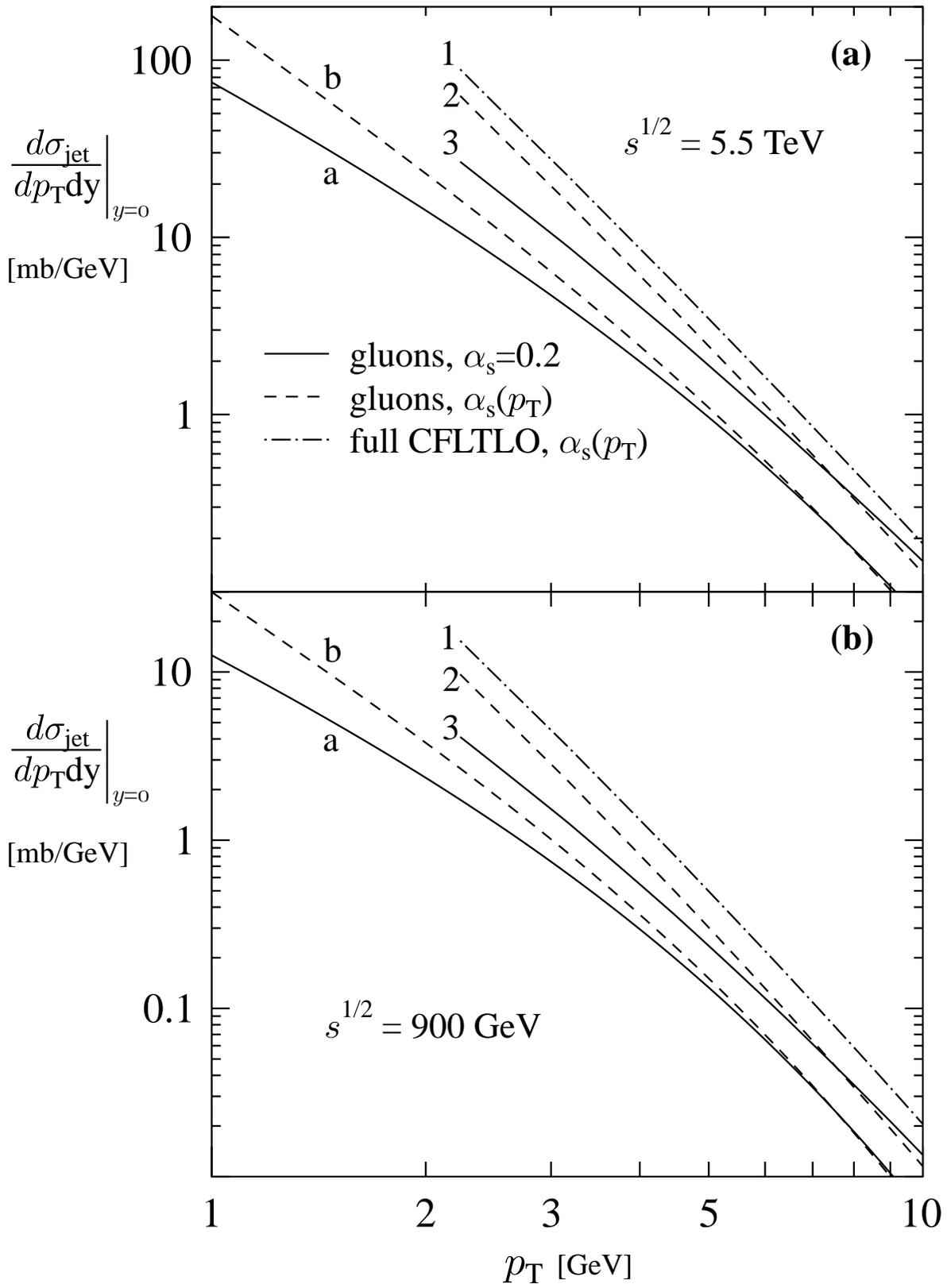
\noindent {\bf Fig.\ 5.}
{\bf a.} Cross sections $d\sigma^{\rm jet}/(dp_{\rm T}dy)$ at $y=0$ as
functions of $p_{\rm T}$ for $\sqrt s = 5.5$ TeV as predicted by
Eq.\ (\ref{MINIJET}). For the BFKL mechanism, two different curves
are shown: the solid line corresponds to having fixed $\alpha_s=0.2$ in
Eq.\ (\ref{MINIJET}), the dashed line corresponds to
$\alpha_s = \alpha_s(p_{\rm T})$. For the collinearly factorized leading
twist lowest order (CFLTLO) computation \cite{EKR}, we show three
different curves: the dash-dotted line is the full computation
with all parton flavours and $\alpha_s(p_{\rm T})$,
the dashed line is the line with gluons only, and the
solid line is obtained with gluons only and $\alpha_s= 0.2$.
To see the deviations from scaling limit (see the text), the log-log plot
is used.
\smallskip

\noindent{\bf b.} The same as in Fig.~5a, but for  $\sqrt s = 900$ GeV.

\end{document}